\documentclass[12pt]{article}
\usepackage[margin=0.8 in]{geometry}    
\usepackage[title,titletoc,toc]{appendix}
\usepackage{authblk}
\usepackage{hyperref}     
 \usepackage{color}
 \usepackage[most]{tcolorbox}
 \tcbset{colback=yellow!10!white, colframe=red!50!black, 
        highlight math style= {enhanced, 
            colframe=red,colback=red!10!white,boxsep=0pt}
        }        
\usepackage{graphicx}
\usepackage{amssymb}
\usepackage{amsmath}
\usepackage{amsfonts}
\usepackage{epstopdf}

\newcommand{\p}{\partial}

\newcommand{\ppy}{\frac{\partial }{\partial y}}

\newcommand{\ppxs}{\frac{\partial^2 }{\partial x^2}}
\newcommand{\ppys}{\frac{\partial ^2}{\partial y^2}}

\newcommand{\lm}{\Lambda}
\newcommand{\hf}{{1\over 2}}
\newcommand{\be}{\begin{equation}}
\newcommand{\br}{\begin{eqnarray}}
\newcommand{\er}{\end{eqnarray}}
\newcommand{\ee}{\end{equation}}
\newcommand{\bt}{\begin{tabular}}
\newcommand{\et}{\end{tabular}}
\newcommand{\bc}{\begin{tcolorbox}}
\newcommand{\ec}{\end{tcolorbox}}

\newcommand{\dd}{\delta}

\newcommand{\Dp}{\frac{d^Dp}{(2\pi)^D}}

\newcommand{\eps}{\epsilon}

\title{Aspects of the map from Exact RG to Holographic RG in AdS and dS}

\author[1, 2]{Pavan Dharanipragada \thanks{\href{mailto:pavand@imsc.res.in}{pavand@imsc.res.in}}}
\author[3]{Semanti Dutta \thanks{\href{mailto:semantidutta@iisc.ac.in}{semantidutta@iisc.ac.in}}}
\author[1,2]{B. Sathiapalan \thanks{\href{mailto:bala@imsc.res.in}{bala@imsc.res.in}}}

\affil[1]{Institute of Mathematical Sciences,CIT Campus, Tharamani, Chennai 600113, India}
\affil[2]{Homi Bhabha National Institute\\Training School Complex, Anushakti Nagar, Mumbai 400085, India}
\affil[3]{Centre for High Energy Physics, Indian Institute of Science, C.V. Raman Avenue, Bangalore 560012, India}

\vspace{0.2 in}

\begin{document}

\hspace{12cm} IMSc/2023/02
{\let\newpage\relax\maketitle}

\maketitle
\begin{abstract}
 In earlier work the evolution operator for the exact RG equation was mapped to a field theory in Euclidean AdS. This gives a simple way of understanding AdS/CFT. We explore aspects of this map by studying a simple example of a Schroedinger  equation for a free particle with time dependent mass. This is an analytic continuation of an ERG like equation. We show for instance that it can be mapped to a harmonic oscillator. We show that the same techniques can lead to an understanding of dS/CFT too.

\end{abstract}
\tableofcontents
\newpage

\section{Introduction}

It has been recognized from the early days of the AdS/CFT correspondence \cite{Maldacena,Polyakov,Witten1,Witten2} that the radial coordinate of the AdS space behaves like a scale for the boundary field theory. This observation follows directly from the form of the AdS metric in Poincare coordinates:
\be  \label{ads}
ds^2 = R^2 \frac{dz^2+dx^\mu dx_\mu}{z^2}
\ee 
This leads naturally to the idea of the ``Holographic" renormalization group: If the AdS/CFT conjecture is correct then radial evolution in the bulk must correspond to RG evolution in the boundary theory [\cite{Akhmedov}-\cite{Meloa:2019}]. 

In \cite{Sathiapalan:2017,Sathiapalan:2019,Sathiapalan:2020} a  mathematically precise connection was made between the exact RG (ERG) equation of a boundary theory and holographic RG equations of a bulk theory in Euclidean AdS (EAdS) space. It was shown that the ERG evolution operator of the boundary theory can be mapped by a field redefinition to a functional integral of a field theory in the bulk  AdS space. This guarantees the existence of an EAdS bulk dual of a boundary CFT without invoking  the AdS/CFT conjecture \footnote{There is still the open question of the locality properties of interaction terms in this bulk field theory. For the case of the $O(N)$ model some aspects of this issue were discussed in \cite{Sathiapalan:2020}.}

Given that the crucial ingredient in this connection with ERG is the form of the metric \eqref{ads} with the factor $z^2$ in the denominator, one is naturally led to ask if similar mappings can be done for the dS metric
\be  \label{ds}
ds^2 = L^2 \frac{-d\eta^2+dx^\mu dx_\mu}{\eta ^2}
\ee 
It too has a scaling form. The difference is that the scale is a time like coordinate - so RG evolution seems to be related to a real time evolution. In fact this metric is related to the EAdS metric by an analytic continuation: $i\eta=z,~iL=R$. Thus real time evolution should be related to RG evolution by analytic continuation. These points have been discussed in many of the early papers on de Sitter holography [\cite{Witten:2001}-\cite{vanderSchaar:2003sz}], (see also \cite{Nastase:2019rsn} for more recent work and further references.)

This paper is an attempt to address the question of whether the mapping of \cite{Sathiapalan:2017} can be generalised to include  for instance dS-CFT. One is also led to explore other kinds of mapping in an effort to understand the nature of this map better. In \cite{Sathiapalan:2017} the map was first introduced in the case of $0$-dimensional field theory in the boundary, which gave a one dimensional bulk field theory or equivalently a point particle quantum mechanical system. In this paper therefore we start by exploring maps for point particle quantum mechanical systems. In Section \ref{freeparticle} we show that the dynamics of a free particle with a time dependent mass can be mapped to a harmonic oscillator. The Euclidean version of this is relevant for the ERG equation. In Section \ref{ERG_dS} the case of mapping a field theory ERG equation to  de Sitter space is considered by starting with the analytically continued form. This complements the discussion of earlier papers where dS-CFT is described as an analytic continuation of EAdS-CFT.
In Section \ref{semiclassicalSolution} we give some examples of two point functions obtained using the techniques of \cite{Sathiapalan:2017} being analytically continued to dS space. Section \ref{conclusion} contains a summary and conclusions.

\section{Mapping Free Particle with Time Dependent Mass to a Harmonic Oscillator}
\label{freeparticle}
In this section we reconsider the construction of \cite{Sathiapalan:2017} where the action for a free field theory in $D+1$ dimension with a non standard  kinetic term was mapped to a free field in $AdS_{D+1}$. When $D=0$ this is just a particle: we will map a free particle with time dependent mass to  a harmonic oscillator. 

\subsection{Mapping Actions} \label{mapping}

\subsubsection{Lorentzian Case}

Consider the following action. It defines an evolution operator for free particle (with time dependent mass) wave function.

\be  \label{A1}
S= \hf \int_{t_i}^{t_f} dt~ M(t) \dot x^2
\ee

\be  \label{A2}
\Psi(x_,t) =\int dx_i \int_{\begin{array}{ccc}
x(t_i)&=&x_i\\
x(t)&=&x
\end{array}} {\cal D}x ~e^{i\hf \int _{t_i}^t M(t') \dot x^2 dt'} \Psi(x_i,t_i)
\ee

Let $x(t)=f(t)y(t)$ with $f^2(t) = \frac{1}{M(t)}$.
Substitute this in \eqref{A1}.
\[
S=\hf \int dt~ ( \dot y^2+ (\frac{\dot f}{f})^2 y^2 + 2 \frac{\dot f}{ f} \dot y y)
\]
\[
=\hf \int dt~ [ \dot y^2+ (\frac{d \ln f}{dt})^2 y^2 -(\frac{d^2}{dt^2} \ln f ) y^2] +\hf\int dt~\frac{d}{dt}(\frac{d \ln f}{dt}y^2)
\]
Thus, upto the boundary term, the action is
\be
S=
\hf \int dt~ [ \dot y^2+  e^{\ln f}(\frac{d^2}{dt^2} e^{-\ln f }) y^2] 
\ee
Now choose
\be   \label{A3.5}
e^{\ln f}(\frac{d^2}{dt^2} e^{-\ln f })=-\omega_0^2
\ee
and we get

\be \label{A3.6}
\bar S=
\hf \int dt~ [ \dot y^2 - \omega_0^2 y^2] 
\ee

which is the  action for a harmonic oscillator. And we define $\bar \Psi$ by absorbing the contribution from the boundary term:

\be  \label{A3.7}
\underbrace{e^{-\hf i \frac{d\ln f(t)}{dt}y^2(t)} \Psi(f(t)y,t)}_{\bar \Psi(y,t)} =\int dy_i \int_{\begin{array}{ccc}
y(t_i)&=&y_i\\
y(t)&=&y
\end{array}} {\cal D}y ~e^{i\hf \int _{t_i}^t [\dot y^2 -\omega_0 ^2 y^2]dt'} \underbrace{e^{-\hf i \frac{d\ln f(t_i)}{dt}y^2(t_i)}\Psi(f(t_i)y_i,t_i)}_{\bar \Psi(y_i,t_i)}
\ee

$\bar S$ thus defines an evolution operator for the harmonic oscillator wave function $\bar \Psi$.
$f$ satisfies
\be   \label{A5}
\frac{d^2}{dt^2}\frac{1}{f} = -\omega_0^2 \frac{1}{f}
\ee
$y$ obeys the same equation.

Thus we can take 
\be  \label{A6}
\frac{1}{f}= a~ cos~ \omega_0(t-t_0)
\ee
which requires
\[
M(t)= a^2 cos^2\omega_0 (t-t_0)
\]

Note that one can do more general cases if one is willing to reparametrize time \cite{Padmanabhan:2017,Ramos}. Thus let
\be
d\tau =\frac{dt}{Mf^2}
\ee


Then one gets \eqref{A3.6}, \eqref{A5} and \eqref{A6}  with $\tau$ replacing $t$. In terms of $t$, \eqref{A5} becomes
\be
\frac{d}{dt}(M\dot f) = \frac{\omega_0^2}{Mf^3}
\ee

Very interestingly, as pointed out in \cite{Padmanabhan:2017},
it is clear from \eqref{A3.6} that the energy of the harmonic oscillator given by 
\[
E= \hf (\dot y^2 + \omega_0^2y^2)
\]
is a conerved quantity. In terms of the original variables this is
\[
E= \hf (  (\frac{\dot x f- x\dot f }{f^2})^2 +\omega_0^2(\frac{x}{f})^2)
\]
These are known as Ermakov-Lewis invariants - see \cite{Padmanabhan:2017} for references to the literature on these invariants - and we see a nice interpretation for them.

\subsubsection{Euclidean Case}
In the Euclidean case the functional integral is
\be  \label{B1}
\Psi(x_,\tau) =\int dx_i \int_{\begin{array}{ccc}
x(\tau_i)&=&x_i\\
x(\tau)&=&x
\end{array}} {\cal D}x ~e^{-\hf \int _{\tau_i}^\tau M(\tau') \dot x^2 d\tau'} \Psi(x_i,\tau_i)
\ee

$\Psi$ in this case is not a wave function. It was shown in \cite{Sathiapalan:2017} that the evolution operator for a $D$-dimensional Euclidean field theory is of this form if we take $M_E(\tau)=-\frac{1}{\dot G(\tau)}$ and $D=0$.  In this case $\Psi$ can be taken to be $e^{-{\cal H}[x_i,\tau_i]}$ where ${\cal H}$ is a Hamiltonian or Euclideanized action. Alternatively (depending on what $M_E(\tau)$ is) it can also be $e^{W[J]}$ - a generating functional or partition function.

Setting $x=fy$ with $f^2= \frac{1}{M_E(\tau)}$, one goes through the same manipulations but replacing \eqref{A3.5} by
\be
e^{\ln f}(\frac{d^2}{d\tau^2} e^{-\ln f })=+\omega_0^2
\ee
 and \eqref{A3.6},\eqref{A3.7} and \eqref{A5} are replaced by
 
\be  \label{B4}
\bar S=
\hf \int d\tau~ [ \dot y^2 +\omega_0^2 y^2]
\ee

\be  \label{B4.7}
\bar \Psi(y,\tau) =\int dy_i \int_{\begin{array}{ccc}
y(\tau_i)&=&y_i\\
y(\tau)&=&y
\end{array}} {\cal D}y ~e^{-\hf \int _{\tau_i}^\tau [\dot y^2 +\omega_0 ^2 y^2]d\tau'} \bar\Psi(y_i,\tau_i)
\ee

and
\be   \label{B5}
\frac{d^2}{d\tau^2}\frac{1}{f} = \omega_0^2 \frac{1}{f}
\ee

The solutions are of the form
\be
f= A ~sech ~\omega_0(\tau-\tau_0)
\ee
which means $M_E(\tau)= \frac{1}{A^2} cosh^2\omega_0(\tau-\tau_0)$.

\eqref{B4.7} has a $\tau$ independent action. In this case there are well known physical interpretations for the Euclidean theory. The evolution operator, $K(y,\tau;y_i,0)$, where
\be
K(y,\tau;y_i,0)=\int_{\begin{array}{ccc}
y(0)&=&y_i\\
y(\tau)&=&y
\end{array}} {\cal D}y ~e^{-\hf \int _{0}^\tau [\dot y^2 +\omega_0 ^2 y^2]d\tau'}
\ee
is the density operator of a QM harmonic oscillator in equilibrium at temperature specified by $\beta = \tau$.

Less well known is that the evolution operator of the Fokker-Planck equation in stochastic quantization can be written in the form given in \eqref{B4.7}. $\bar\Psi$ is then related to the probability function (see, for instance, \cite{Damgaard:1987} for a nice discussion).

In the next section we discuss the mappings directly for the Schroedinger equation, rather than its evolution operator.



%

\subsection{Mapping Schrodinger Equations}
\subsubsection{Lorentzian}  \label{LSE}

Let us consider the same mapping from the point of view of the Schroedinger equation for the free particle wave function.

Schrodinger's equation for the free particle is

\be  \label{C1}
i\frac{\p \Psi(x,t)}{\p t}=-\frac{1}{2M(t)}\frac{\p^2 \Psi(x,t)}{\p x^2}
\ee
$\Psi$ given by \eqref{A2} obeys this equation.

We make a coordinate transformation and a wave function redefinition. Both can be understood as canonical transformations \cite{Anderson:1992}.

Let $x= f(t)y$ with $f^2=\frac{1}{M(t)}$.  We take $f,M$ to be dimensionless. We treat this as a $0+1$ dimensional field theory where $x$ has the canonical dimension of $-\hf$. So $x=L^{\hf}X$ would define a dimensionless $X$. $L$ is some length scale.
\[
\frac{\p \Psi(x,t)}{\p t}= \frac{\p \Psi(f(t)y,t)}{\p t} - \frac{\dot fy}{f}\frac{\p \Psi(f(t)y,t)}{\p  y}
\]

Let
\[
\Psi (f(t)y,t)= e^{-\hf \alpha y^2}\bar \Psi(y,t)
\]
\[
\frac{\p \Psi}{\p t} = e^{-\hf \alpha y^2}(-\hf \dot \alpha y^2 + \frac{\p}{\p t})\bar \Psi(y,t)
\]
\[
-i \frac{\dot fy}{f}\frac{\p \Psi(f(t)y,t)}{\p  y}
=ie^{-\hf \alpha y^2}(\alpha \frac{\dot f}{f} y^2 - \frac{\dot f}{f}y \ppy)\bar \Psi (y,t)
\]
\[
\frac{1}{M}\hf \ppxs \Psi=\hf \ppys e^{-\hf \alpha y^2}\bar \Psi=( \hf e^{-\hf \alpha y^2}(\alpha^2 y^2 - 2\alpha y \ppy -\alpha + \ppys)\bar \Psi)
\]

Collecting all the terms one finds that \eqref{C1} becomes:
\be   \label{C2}
i\frac{\p \bar\Psi}{\p t}= (\hf i \dot \alpha -i\alpha \frac{\dot f}{f} -\hf \alpha^2)y^2\bar \Psi +(i\frac{\dot f}{f} y\ppy + \alpha y\ppy) \bar \Psi +\hf \alpha \Psi - \frac{1}{2}\ppys \bar \Psi
\ee

We choose $\alpha = -i\frac{\dot f}{f}$ to get rid of the second term on the RHS. We get
\[
i\frac{\p \bar\Psi}{\p t}=
[(\hf \frac{d^2 \ln f}{dt^2} -\hf (\frac{d\ln f}{dt})^2) y^2 +\hf \alpha -\frac{1}{2}\ppys ]\bar \Psi
\]
As before it can be rewritten as
\be   \label{C3}
i\frac{\p \bar \Psi}{\p t}=\hf [-e^{\ln f}(\frac{d^2}{dt^2} e^{-\ln f})y^2-\ppys +{\alpha}]\bar \Psi
\ee

Set
\[
\frac{d^2}{dt^2} \frac{1}{f}= -\omega_0^2\frac{1}{f}
\]
 again as before to get
\be   \label{C4}
i\frac{\p \bar\Psi}{\p t}= \frac{1}{2}[-\ppys +\omega_0^2 y^2 + \alpha]\bar \Psi
\ee 

The term $\hf \alpha$ generates a scale transformation
$e^{- \hf\ln \frac{f(t)}{f(t_i)}}$ for $\bar \Psi$.


\subsubsection{Euclidean} \label{ESE}
The Euclidean version is
\be  \label{C5}
\frac{\p \Psi(x,\tau)}{\p \tau}=\frac{1}{2M_E(\tau)}\frac{\p^2 \Psi(x,\tau)}{\p x^2}
\ee

As mentioned above, this is of the form of a Polchinski ERG equation (with $\frac{1}{2M_E(\tau)}=-\dot G(\tau)$)  for 
${\cal H}$ defined by $\Psi\equiv e^{-{\cal H}}$.
Going through the same steps one finds, with $f^2= \frac{1}{M_E(\tau)}$,
\be   \label{C5.5}
\frac{\p \bar\Psi}{\p \tau}= (\hf  \dot \alpha -\alpha \frac{\dot f}{f} +\hf \alpha^2)y^2\bar \Psi +(\frac{\dot f}{f} y\ppy - \alpha y\ppy) \bar \Psi -\hf \alpha \Psi + \frac{1}{2}\ppys \bar \Psi
\ee

the condition
$\alpha = \frac{\dot f}{f}$ and the equation becomes

\be   \label{C6}
\frac{\p \bar \Psi}{\p t}=\frac{1}{2} [-\underbrace{e^{\ln f}(\frac{d^2}{dt^2} e^{-\ln f})}_{=~\omega_0^2}y^2+\ppys-\alpha]\bar \Psi
\ee
Thus
\be   \label{C7}
\frac{\p \bar \Psi}{\p \tau}=\frac{1}{2} [\ppys-\omega_0^2y^2-\alpha]\bar \Psi
\ee

And $f$ obeys
\be
\frac{d^2}{dt^2}\frac{1}{f}= \omega_0^2 \frac{1}{f}
\ee
This is a Euclidean harmonic oscillator equation. Various physical interpretations of this equation were given in the last section. The term $\alpha$ in \eqref{C7} provides a multiplicative scaling
$e^{-\hf \int_{t_i}^tdt'~\p_{t'}\ln f}=(\frac{f(t_i)}{f(t)})^\hf$ of $\bar \Psi$.

\subsubsection{Analytic Continuation}
If we set $it=\tau$, \eqref{C1} becomes \eqref{C5} provided $M(-i\tau)=M_E(\tau)$. Similarly \eqref{C4} becomes \eqref{C7}. Note that in \eqref{C4} 
$\alpha = -i\frac{\dot f}{f}$. This analytically continues to $\frac{\dot f}{f}$ as required.

\subsection{Semiclassical Treatment}

Most of the AdS/CFT calculations invoke large N to do a semiclassical treatment of the bulk theory- one can evaluate boundary Green's function. The 
analysis in \cite{Sathiapalan:2017,Sathiapalan:2020} did this for the ERG treatment - the evolution of the Wilson action/Generating functional were calculated.  In \cite{Maldacena:2002} a semiclassical treatment was used to obtain the ground state wave function in dS space. 

For completeness we do the same for the simple systems discussed in this paper. This illustrates the connection between ERG  and  dS.

\subsubsection{Using Harmonic Oscillator Formulation}

Since 
\be  \label{C8}
\Psi(x,t) =\int dx_i \int_{\begin{array}{ccc}
x(t_i)&=&x_i\\
x(t)&=&x
\end{array}} {\cal D}x ~e^{i \int _{t_i}^t L (x(t'),\dot x(t'),t')dt'} \Psi(x_i,t_i)
\ee

solves Schroedinger's equation. For the Harmonic Oscillator
\be 
L= \hf (\dot x^2  - \omega_0x^2)
\ee
for the Lorentzian version.

One can evaluate the path integral semiclassically by plugging in a classical solution with some regular boundary condition.
We choose $x=0$ at $t=-\infty$. The initial state wave function is thus a delta function. Classical solution of the EOM is of the form
\[
x(t) = a e^{-i\omega_0 t} + a^* e^{i\omega_0 t}
\]
Since $a$ should annihilate the vacuum state in the far past we would like the solution to look like
\[
x(t) \to e^{i\omega_0 t} 
\]
in order to ensure that we are in the ground state.

\be  \label{C9}
x(t)= x_f e^{-i\omega_0(t_f-t)}
\ee
At $t=-\infty$ we assume that the solution vanishes.  This is justified by an infinitesimal rotation $t \rightarrow t+i\eps t$.
Evaluated on this solution, the action becomes
\[
S_{classical}=~\hf x(t)\dot x(t)|^{t_f}_{-\infty}
\]

We get
\be \label{C11}
S_{classical}=\hf i\omega_0 x_f^2
\ee

Plugging \eqref{C9} into \eqref{C8} we obtain
\be 
\Psi(x_f) \approx e^{-\hf \omega_0x_f^2}
\ee 

If we repeat this for the free field in dS space we get the ground state wave functional \cite{Maldacena:2002}.

\subsubsection{Using ERG formulation}

For the Euclidean version, we set $it=\tau$ and  we write
\be  \label{C8.5}
\Psi(x,\tau) =\int dx_i \int_{\begin{array}{ccc}
x(\tau_i)&=&x_i\\
x(\tau)&=&x
\end{array}} {\cal D}x ~e^{- \int _{\tau_i}^\tau L_E (x(\tau'),\dot x(\tau'),\tau')d\tau'} \Psi(x_i,\tau_i)
\ee
It is well known that if one does the semiclassical analysis for the Euclidean  case with  general
boundary condition one recovers the thermal density matrix. This is for the time independent Hamiltonian - such as the harmonic oscillator. 
We will not do this here. Instead we proceed directly to the  ERG interpretation of the calculation. Here the Hamiltonian is time dependent. 
%
%
%
%
%
%
%
%
%
%
%
%
In \cite{Sathiapalan:2017} the analysis given below was applied to $W[J]$. We repeat it here for the Wilson action.

\vspace{0.1 in}

Our starting action in this case is (Note $\dot G<0$):
\be
S=-\hf \int _{\tau_i}^{\tau_f} \frac{ \dot x ^2}{\dot G}
\ee
EOM is given by,
\[
\p_\tau (\frac{\dot x}{\dot G})=0
\]
\[
\frac{\dot x}{\dot G}=b \implies x=bG+c
\]
We choose $G$ so that it vanishes at $\tau=\infty$ .

\vspace{0.1 in}

For the Euclidean Harmonic oscillator case $G$ has then to be
\[
G=-\frac{1}{\omega_0}(tanh~\omega(\tau-\tau_i)-1)
\]
Also $x\to 0$ as $\tau\to \infty$. So $c=0$.

\be
x=bG
\ee
\[
x(\tau)=- \frac{b}{\omega_0 }(tanh~\omega(\tau-\tau_i)-1)
\]

On shell 
\[
S=-\hf \int_{\tau_i}^{\tau_f} d\tau~\frac{d}{d\tau}(\frac{x\dot x}{G})
\]
\[=
\hf(x(\tau_f)-x(\tau_i))b=\hf[\frac{x(\tau_f)x(\tau_f)}{G(\tau_f)}-\frac{x(\tau_i)x(\tau_i)}{G(\tau_i)}]
\]

If we add this change to the initial Wilson action
$\hf\frac{x(\tau_i)x(\tau_i)}{G(\tau_i)}$ we get the final Wilson action
\[
{\cal H}_f=\hf\frac{x(\tau_f)x(\tau_f)}{G(\tau_f)}
\]

If, for instance,  we are interested in evaluating ${\cal H}$ semiclassically at $\tau=\tau_i$.
\[
x(\tau_i)=\frac{b}{\omega_0 } \implies b=x(\tau_i)\omega_0
\]
\[
x(\tau)= -x(0)(tanh~\omega(\tau-\tau_i)-1)
\]
\[
\dot x(\tau) =-x(0)\omega_0 sech^2 \omega_0(\tau-\tau_i)
\]
The classical action is
\[
S_{classical} = \hf \omega_0x(\tau_i)^2
\]
Thus since $G(\tau_i)=\frac{1}{\omega_0}$, $\cal H$ evaluated semiclassically is:
\be
{\cal H}[x,\tau_i]\approx \hf\omega_0x(\tau_i)^2
\ee

\vspace{0.1 in}

Then \[
\Psi=e^{-{\cal H}[x,\tau_i]}=e^{-\omega_0 x(\tau_i)^2}
\]
which coincides with the ground state wave function of the harmonic oscillator. This is essentially the
Hartle Hawking prescription \cite{Hartle:1983}. This also motivates the 
dS-CFT correspondence statement \cite{Witten:2001,Strominger:2001,Maldacena:2002} that $\Psi_{dS}=Z_{CFT}$

This concludes the discussion of the mapping of ERG equation to a Euclidean harmonic oscillator. In higher dimensions this gives   free field theory in flat space. We now return to the case of interest, namely dS space.

\section{ERG to field theory in dS}
\label{ERG_dS}

We first map the system to Euclidean AdS. Then analytically continue and obtain dS results. Alternatively, one can analytically continue the ERG equation to the Schroedinger equation (when $D=0$ this is  a free particle with a time dependent mass) and then map to de Sitter space. This is all exactly as was done for the harmonic oscillator.

\subsection{Analytic Continuation}

The EAdS metric in Poincare coordinates is
\be  \label{adm}
ds^2= R^2[\frac{dx_idx^i + dz^2}{z^2}]
\ee

The dS metric in Poincare coordinates is:
\be  \label{dm}
ds^2= L^2[\frac{dx_idx^i - d\eta^2}{\eta^2}]
\ee

The metrics are related by analytic continuation:

\[ i\eta =z,~~~iL=R\]

\subsubsection{Analytic Continuation of the Action}
The action generically is
\be
S=-\hf\int d^{D+1}x \sqrt{g} [g^{\mu \nu}\p_\mu \phi \p_\nu\phi +m^2 \phi^2]
\ee

\paragraph{ de Sitter}

In this case we write $\sqrt{-g}$ since $g$ is negative: $g = -(\frac{L^2}{\eta^2})^{D+1}$. Also
$ g^{00}=  -\frac{\eta^2}{L^2}$ and $g^{ij}=\dd^{ij}\frac{\eta^2}{L^2}$.

Thus
\be
S_{dS}=\int d^Dx \int_{0}^\infty d\eta ~(\frac{L}{\eta})^{D+1}[\frac{\eta^2}{L^2} \p_\eta \phi \p_\eta \phi -\frac{\eta^2}{L^2}\p_i\phi \p_i \phi - m^2 \phi^2]
\ee
In momentum space:
\be
S_{dS}=\int \Dp \int_{0}^\infty d\eta ~(\frac{L}{\eta})^{D+1}[\frac{\eta^2}{L^2} \p_\eta \phi(p) \p_\eta \phi(-p) -(\frac{\eta^2}{L^2}p^2+m^2)\phi(p) \phi(-p) ]
\ee

The functional integral description of the quantum mechanical evolution operator for the wave functional of the fields in dS space-time is
\be  \label{zd}
\bar \Psi[\phi(p),t] =\int d\phi_i(p) \int_{\begin{array}{ccc}
\phi(p,t_i)&=&\phi_i(p)\\
\phi(p,t)&=&\phi(p)
\end{array}} {\cal D}\phi(p,t) ~e^{i\hf \int _{t_i}^t [\dot \phi(p,t')^2 -\omega_0 ^2 \phi(p,t')^2]dt'} \bar\Psi[\phi_i(p),t_i]
\ee

\paragraph{ Euclidean Anti de Sitter}

$g = (\frac{R^2}{z^2})^{D+1}$. Also
$ g^{00}=  \frac{z^2}{R^2}$ and $g^{ij}=\dd^{ij}\frac{z^2}{R^2}$.

\be
S_{EAdS}=\int d^Dx \int_0^\infty dz ~(\frac{R}{z})^{D+1}[\frac{z^2}{R^2} \p_z \phi \p_z \phi +\frac{z^2}{R^2}\p_i\phi \p_i \phi + m^2 \phi^2]
\ee
In momentum space
\be
S_{EAdS}=\int \Dp \int_0^\infty dz ~(\frac{R}{z})^{D+1}[\frac{z^2}{R^2} \p_z \phi(p) \p_z \phi(-p) +(\frac{z^2}{R^2}p^2+m^2)\phi (p) \phi(-p) ]
\ee

If we set $i\eta =z$ and $iL=R$ we see that the functional integral \eqref{zd} becomes
\be  \label{zad}
\bar \Psi[\phi(p),t] =\int d\phi_i(p) \int_{\begin{array}{ccc}
\phi(p,t_i)&=&\phi_i(p)\\
\phi(p,t)&=&\phi(p)
\end{array}} {\cal D}\phi(p,t) ~e^{-\hf \int _{t_i}^t [\dot \phi(p,t')^2 +\omega_0 ^2 \phi(p,t')^2]dt'} \bar\Psi[\phi_i(p),t_i]
\ee
In holograhic RG this is interpreted as a Euclidean functional integral giving the evolution in the radial direction.  $\bar \Psi$ is to be interpreted as $e^{-S_I[\phi(p),t]}$ where $S_I$ is the Wilson action. It was shown in \cite{Sathiapalan:2017} (see below) that this can be obtained by mapping an ERG evolution operator.

The dS functional integral \eqref{zd}  above is thus an analytically  continued version of this. 

\subsection{Mapping}

\subsubsection{Mapping from Quantum Mechanics}

Let us go back to Section \eqref{mapping} and consider the mapping from the Quantum Mechanics of a free particle with time dependent mass. We think of it as a $0+1$ dimensional field theory. $M(t)$ is taken to be dimensionless and $x$ has canonical dimensions of $-\hf$. 

\be  \label{D1}
S= \hf \int dt~  M(t)\dot x^2
\ee
(In the ERG version $M(t)=\frac{1}{\dot G}$)

The path integral is 
\be \label{D2}
\int {\cal D}x ~ e^{i S}
\ee

As before $x(t)=f(t)y(t)$ with $f^2(t) = \frac{1}{M(t)}$.
Substitute this in \eqref{D1} and go through the same steps to obtain:
\be
S=
\hf \int dt~ [ \dot y^2+  e^{\ln f}(\frac{d^2}{dt^2} e^{-\ln f }) y^2] 
\ee
Now choose
\be   \label{D3.5}
e^{\ln f}(\frac{d^2}{dt^2} e^{-\ln f })= -(\frac{\eta^2}{L^2} p^2 +m^2)
\ee

where $\eta = L e^{\frac{t}{L}}$. to obtain $S_{dS}$
\[
S_{dS} =
\hf \int dt~ [ \dot y^2 -(\frac{\eta^2}{L^2} p^2 +m^2)  y^2] 
\]
\be	\label{D4}
=\hf\int d\eta ~(\frac{L}{\eta})[\frac{\eta^2}{L^2} \p_\eta y \p_\eta y -(\frac{\eta^2}{L^2}p^2+m^2)y^2 ]
\ee

$p,m$ here are just some parameters. When $D>0$ they will stand for momentum and mass of the field respectively. So starting from a free particle with time dependent mass we obtain the free field action in de Sitter space $dS_{D+1}$ with $D=0$.

\paragraph{Schroedinger Equation:}

\be  \label{D5}
i\frac{\p \Psi(x,t)}{\p t}=-\frac{1}{2M(t)}\frac{\p^2 \Psi(x,t)}{\p x^2}
\ee

Using the same mapping as in Section \eqref{LSE}, $x=fy$ 
\[
\Psi (f(t)y,t)= e^{-\hf \alpha y^2}\bar \Psi(y,t)
\]
with $\alpha = -i\frac{\dot f}{f}$ one obtains
\[
i\frac{\p \bar\Psi}{\p t}=
[(\hf \frac{d^2 \ln f}{dt^2} -\hf (\frac{d\ln f}{dt})^2) y^2 +\hf \alpha -\hf \ppys ]\bar \Psi
\]
Using \eqref{D3.5} this becomes
\be	\label{D6}
i\frac{\eta}{L} \frac{\p \bar\Psi}{\p \eta}=
[-\hf \ppys+\hf(\frac{\eta^2}{L^2}p^2 + m^2) y^2 +\hf \alpha  ]\bar \Psi
\ee

If we construct the Schroedinger equation corresponding to the action \eqref{D4} one obtains 
\be  \label{D7}
i\frac{\eta}{L} \frac{\p \bar\Psi}{\p \eta}=
[-\hf \ppys+\hf (\frac{\eta^2}{L^2}p^2 + m^2) y^2   ]\bar \Psi
\ee

which barring the field independent term $\alpha$ 
is exactly the same as \eqref{D6}. This term as we have seen provides an overall field independent scaling for all wave functions. It is a consequence of the ordering ambiguity in going from  classical to quantum treatment. 
\eqref{D7} (or its extension to $D>0$) describes the quantum mechanical time evolution of the matter field wave functional in de Sitter space.

\subsubsection{Mapping from ERG}

\paragraph{ Action}

We now consider the Euclidean version of \eqref{D1}, which is the Polchinski ERG equation. This is what was done in  \cite{Sathiapalan:2017}.  Thus we replace $M(t)$ by $-\frac{1}{\dot G}$.
\be  \label{D8}
S= -\hf \int d\tau~  \frac{\dot x^2}{\dot G}
\ee

The path integral is ($\dot G<0$)
\be \label{D9}
\int {\cal D}x ~ e^{\hf \int d\tau~  \frac{\dot x^2}{\dot G} }
\ee
which can be obtained from  \eqref{D5} by setting $it=\tau$. 
We take $z= Re^{\frac{\tau}{R}}$ If we let
$i\eta =z,~iL=R,~it=\tau$ then this can be obtained from the corresponding Minkowski case.

As before $x(\tau)=f(\tau)y(\tau)$ with $f^2(\tau) = \dot G$.
Substitute this in \eqref{D8} and go through the same steps to obtain:
\be
S=
\hf \int d\tau~ [ \dot y^2+  e^{\ln f}(\frac{d^2}{d\tau^2} e^{-\ln f }) y^2] 
\ee
Now choose
\be   \label{D10}
e^{\ln f}(\frac{d^2}{d\tau^2} e^{-\ln f })= (\frac{z^2}{R^2} p^2 +m^2)
\ee

where $z = R e^{\frac{\tau}{R}}$. to obtain $S_{EAdS}$
\be	\label{D11}
S_{EAdS}=\int dz ~(\frac{R}{z})[\frac{z^2}{R^2} \p_z y \p_z y +(\frac{z^2}{R^2}p^2+m^2)y^2 ]
\ee

\paragraph{ERG Equation}

By analogy with the Schroedinger equation we can see that \eqref{D9} is the evolution operator corresponding to the ERG equation 
\be  \label{D11}
\frac{\p \Psi(x,\tau)}{\p \tau}=-\hf \dot G\frac{\p^2 \Psi(x,\tau)}{\p x^2}
\ee

By the same series of transformations as in the de Sitter case, but using \eqref{D10}, one obtains
\be	\label{D12}
\frac{z}{R} \frac{\p \bar\Psi}{\p z}=
[\hf \ppys -(\frac{z^2}{R^2}p^2 + m^2) y^2 -\hf \alpha ]\bar \Psi
\ee

with $\alpha = \frac{\dot f}{f}$ generating an overall scale transformation for $\bar \Psi$. In the ERG context $\bar \Psi$ represents $e^{W[J]}$ upto a quadratic term. This equation is the holographic RG equation in the AdS/CFT correspondence for an elementary scalar field \cite{Sathiapalan:2017}. 

\subsection{Connections}

Let us summarize the various connections obtained above. 
\begin{itemize}

\item We start with the quantum mechanics of a free particle having a time dependent mass. The Schroedinger equation (SE) for this is \eqref{C1}. Analytical continuation of this equation (generalized to higher dimensions) gives the Polchinski ERG equation \eqref{C5}. 

\item The free particle SE \eqref{C1} can be mapped to a SE for a harmonic oscillator \eqref{C4}. The ERG equation \eqref{C5} can similarly  be mapped to a Euclidean harmonic oscillator \eqref{C7}- analytically continued version  of \eqref{C4}. 

\item The evolution operators for the above equations are defined in terms of path integrals over some actions. The same mapping function $f$ maps the corresponding actions to each other. Thus the evolution operator for the free particle Schroedinger equation is given by the action in \eqref{A1} which is mapped to a harmonic oscillator action \eqref{A3.6}. The analytical continuation of these are the Euclidean ERG evolution operator \eqref{B1}  mapped  to a harmonic oscillator Hamiltonian \eqref{B4.7}. These steps are summarized in the flow diagram in Figure 1.

\item The mapping function $f$ was originally chosen in \cite{Sathiapalan:2017} to map the free particle ERG action \eqref{D8} to an action for free fields in  $EAdS_{0+1}$ given in \eqref{D11}.  The analytical continuation of this problem to real time gives us an action in $dS_{0+1}$ \eqref{D4}. 

\item One can also repeat these steps for the corresponding ``wave" equations. The Polchinski ERG equation for $e^{W[J]}$ gets mapped to an equation in EAdS for $e^{W[J]}$ which is nothing but the holographic RG equations. Analytically continuing this, the Schroedinger equation for a wave functional is mapped to a Schroedinger equation for wave functionals of fields in dS. 

\end{itemize}

These are summarized in the figure below (Fig.2).  The analytic continuation can be done before the map with $f$ is applied or after as shown in the figure. It can be done both for the actions as well as for the equations.

 \begin{figure}[h]
    \centering
    \includegraphics[width=8cm]{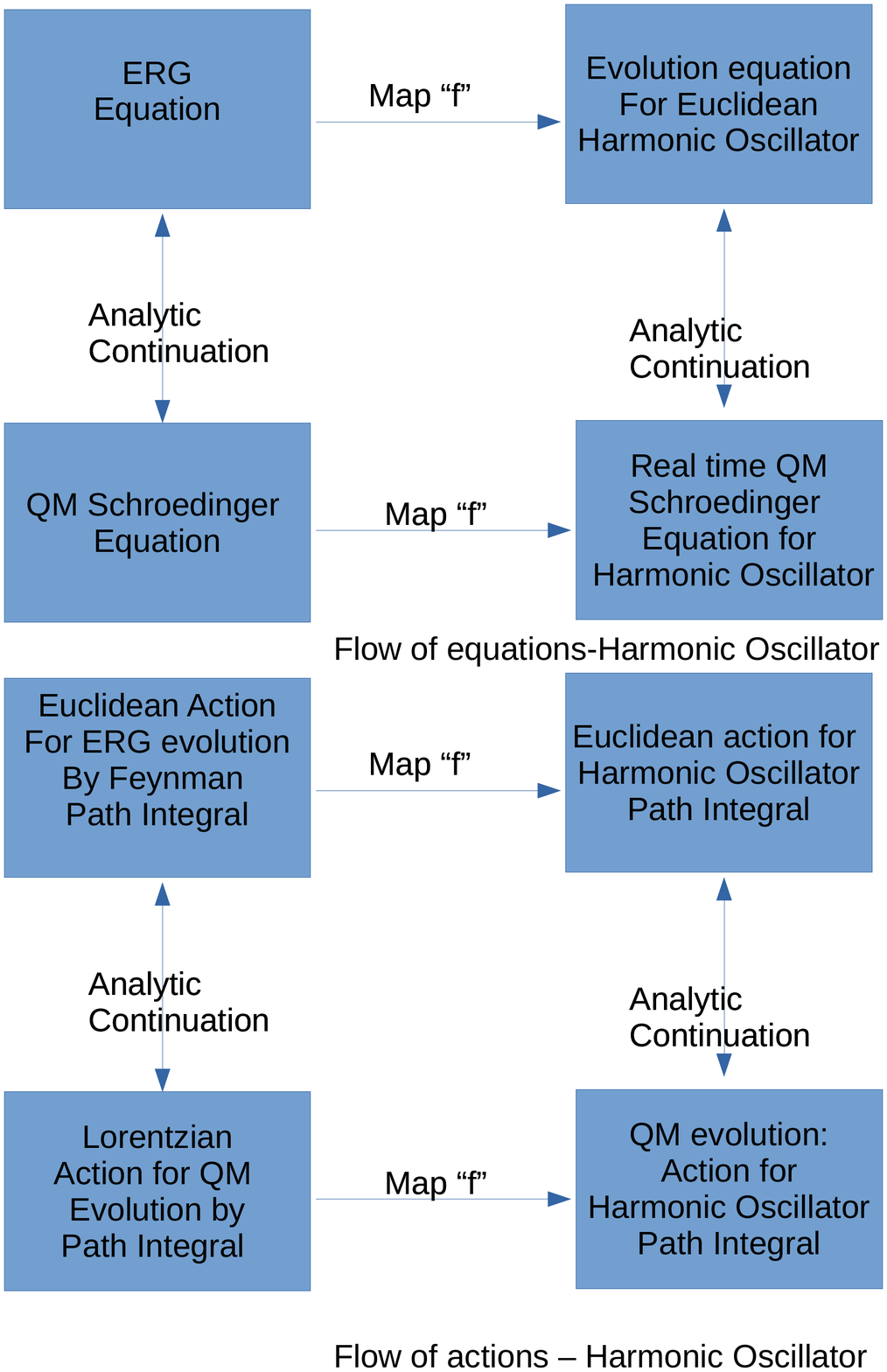}
    \caption{Mapping ERG to Harmonic Oscillator}
    \label{Fig 1}
\end{figure}

\begin{figure}[h]
    \centering
    \includegraphics[width=8cm]{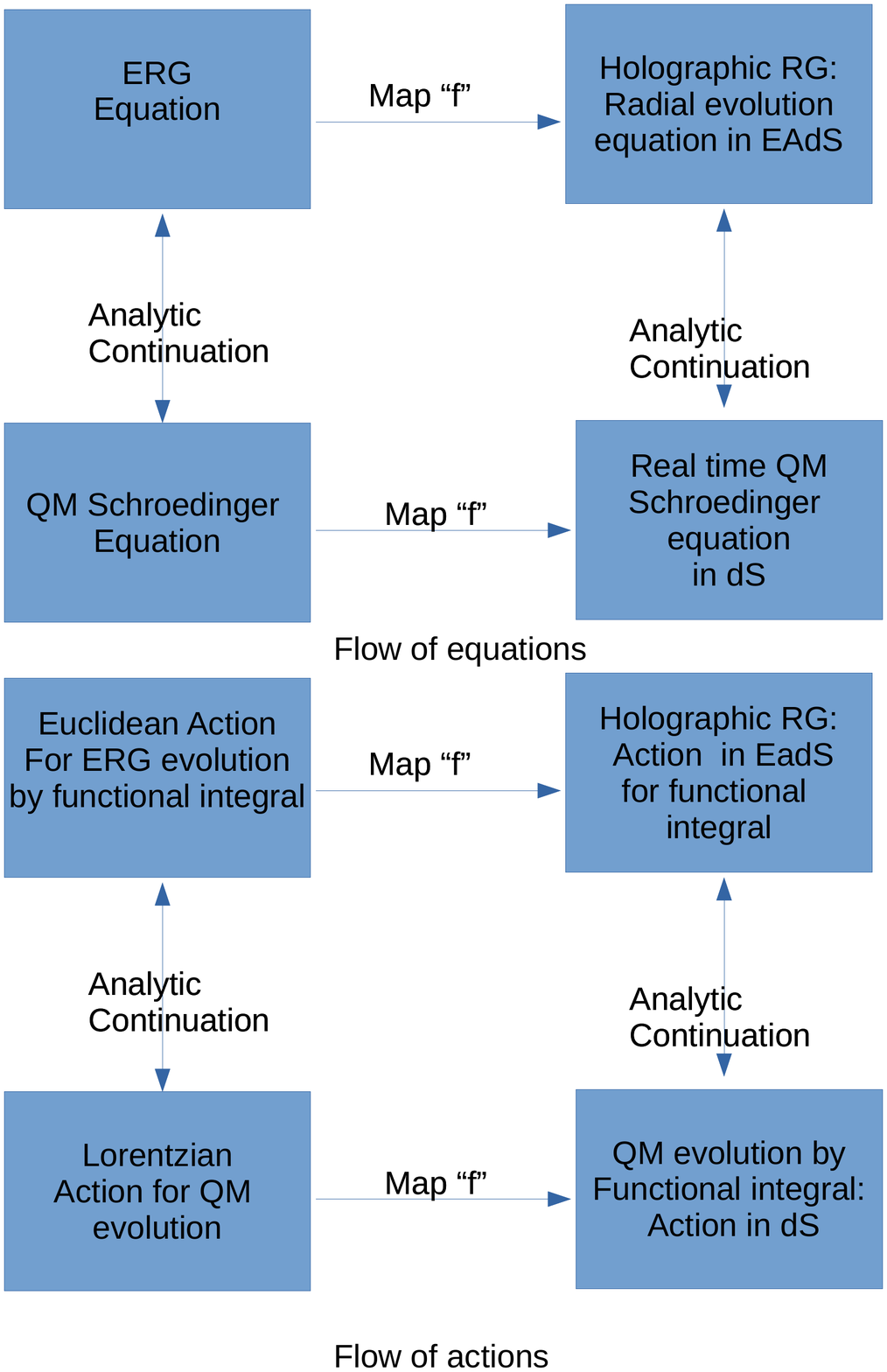}
    \caption{Mapping ERG to Holographic RG}
    \label{Fig 2}
\end{figure}

\subsection{dS-CFT correspondence}

The connections with ERG mentioned above should, if pursued, provide some insights into dS-CFT correspondence. We restrict ourselves to some preliminary observations in this paper.

The idea of dS-CFT correspondence was suggested in
\cite{Witten:2001,Strominger:2001,Maldacena:2002}.
This has been investigated further by many authors, e.g.  \cite{Bousso:2001,Spradlin:2001,Harlow:2011,Das:2013,Anninos:2011a,Anninos:2011b,Anninos:2014}. 

What we see from the above analysis is that considering the {\em relation between the evolution equations},  one can say that

\be  \label{dcft}
\Psi[\phi,J]_{wave-functional~in ~dS} = \{Z[\phi,J]_{CFT}\}_{analytically~continued}
\ee

Thus we see that the dS-CFT correspondence suggested by this analysis is one between an ERG equation for a CFT generating functional and a {\em real time quantum mechanical evolution} of a wave functional in dS space time.

The LHS of \eqref{dcft} is a QM wave functional of fields on a $D$-dimensional spatial slice of a $D+1$ dimensional dS spacetime. The RHS is the analytically continued partition function of a $D$-dimensional Euclidean CFT - more precisely, either $e^{W_\lm[J]}$ or  $e^{-S_{I,\lm}[\phi]}$. The precise statement has to involve some statement of the boundary conditions. In the next section we give a concrete example with boundary conditions specified.

Note that the LHS is a complex probability {\em amplitude}. Expectation values will involve $\Psi^*\Psi$ and were calculated first in \cite{Witten:2001,Strominger:2001,Maldacena:2002}.
 
 One can proceed to ask whether the expectations on the spatial slice calculated using $\Psi^*\Psi$ also correspond to some other Euclidean CFT on the spatial slice.  This was explored further in \cite{Harlow:2011}.  We do not address this question here.

In the next section we give some examples that explicitly illustrate the connection made by \eqref{dcft}. 
\newpage

\section{Obtaining Bulk field from ERG}
\label{semiclassicalSolution}
The ERG formulation stated in this paper starts with  the boundary fields. The evolution operator for this  involves bulk fields but with a non standard action.  When this action is mapped to EAdS action one can interpret the newly mapped field as the EAdS bulk field. This analysis for Euclidean AdS is well defined and has been done in \cite{Sathiapalan:2017,Sathiapalan:2020}. However, this treatment does not have a natural interpretation in the context in dS space. We have elaborated that in this section. 

\subsection*{Bulk scalar field in Euclidean AdS and dS}

There are conceptual barriers if one tries to do similar analysis to map the ERG evolution operator directly to Lorentzian dS. 
First of all, it is not clear as in EAdS whether the function  G(t) a.k.a $f^2(t)=\dot{G}(t)$ is the Green's function of the dual field theory of dS. It has an oscillatory cutoff function. 
Therefore we analytically continue the ERG action to a Lorentzian action first, and then do the mapping.

 The result thus obtained \eqref{dsbulk} matches with the value found in \cite{Das:2013} where the authors have found the bulk field in semicalssical approximation from dS bulk action. For the Lorentzian dS analysis presented here  the RG interpretation is not clearly understood - except as an anlytic continuation.  We have presented it here for sake of completeness.

\paragraph{Euclidean AdS}

The Euclidean action of the ERG evolution operator in momentum space,
\be  \label{erg3}
S=-\hf \int d\tau \int_p ~\frac{\dot \phi^2}{\dot G}
\ee is mapped to 
\be
S_{EAdS}=\int \Dp \int_{\eps_{EAdS}}^\infty dz ~(\frac{R}{z})^{d+1}[\frac{z^2}{R^2} \p_z y^{EAdS}(p) \p_z y^{EAdS}(-p) +(\frac{z^2}{R^2}p^2+m^2)y^{EAdS} (p) y^{EAdS}(-p) ]
\ee

with $z=Re^{\frac{\tau}{R}}$ as described in \cite{Sathiapalan:2017}.
We have rescaled the field as $\phi= f y^{EAdS}$ where $f$ is related to the boundary Green's function G as $f^2= -\left(\frac{z}{R}\right)^{-d} \dot G$.

The constraint on $\frac{1}{f}$ is given by,

\be
\frac{\p}{\p z} \lbrace \left( \frac{z}{R} \right)^{-d+1} \frac{\p}{\p z} \frac{1}{f} \rbrace= \left( \frac{z}{R} \right)^{-d+1} \left(p^2+ \frac{m^2 R^2}{z^2}\right) \frac{1}{f}
\ee

The solutions are $z^{d/2} K_\alpha(pz)$ and $z^{d/2} I_\alpha(pz)$ where $\alpha^2=m^2 R^2+\frac{d^2}{4}$.

So $\frac{1}{f}$ can be taken as,

\be
\frac{1}{f(p,z)}= (z)^{d/2} \left(A K_\alpha (pz)+ B I_\alpha (pz)\right)
\ee

The Green's function is

\be
G(p,z)= \frac{C K_\alpha (pz)+ D I_\alpha(pz)}{A K_\alpha(pz)+ B I_\alpha(pz)}
\ee

The large argument asymptotic form of the Modified Bessel function $I_\alpha(z)$ and $K_\alpha(z)$ are given by,

\[ I_\alpha(z) \sim \frac{e^z}{\sqrt{ 2\pi z}} \left(1+ \mathcal{O}(\frac{1}{z}) \right) ~~ for~~ |arg~z|<\frac{\pi}{2}\]
\[ K_\alpha(z) \sim \sqrt{\frac{\pi}{2z}} e^{-z} \left(1+ \mathcal{O}(\frac{1}{z}) \right) ~~ for~~ |arg~z|<\frac{3\pi}{2}\]

Putting two constraints on G- i)$G(pz \rightarrow \infty)=0$ ii)$G(pz \rightarrow 0)= \gamma_{EAdS}~ p^{-2\alpha}$, we get,

\[ D=0; ~ C(p)=\gamma_{EAdS}~ p^{-\alpha};~ B(p)=-\frac{1}{\gamma_{EAdS}} p^\alpha \]

In semiclassical approximation the bulk field $y^{EAdS}= b_{EAdS}\frac{G}{f}$. If $y^{EAdS}$ satisfies $y_0^{EAdS}$ the bulk field is given by,

\be
y^{EAdS}=y_0^{EAdS} \frac{z^{d/2}}{\eps^{d/2}} \frac{K_\alpha(pz)}{K_\alpha(p\eps)}
\ee

Now let's check by analytic continuation $i\eta=z$ and $iL=R$. First of all, $\alpha$ becomes $\nu$. $\eps$ is replaced by $i\eps$. We get,

\be
y^{EAdS}|_{z= i\eta,~R= iL}= y_0^{EAdS}|_{z=i\eta,~R=iL} \frac{(i\eta)^{d/2}}{(i\eps)^{d/2}} \frac{K_\nu(ip\eta)}{K_\nu(ip\eps)}
\ee

As,

\be
y_0^{EAdS}= b_{EAdS}~\eps_{EAdS}^{d/2} \frac{\gamma_{EAdS}~K_\alpha(p\eps)}{p^\alpha}
\ee

\paragraph{de Sitter}

We would like to do the same analysis as above for the Lorentzian case.

The Lorentzian action obtained from \eqref{erg3} by analytic continuation, in momentum space,
\[
S= -\int dt~\int \Dp \frac{1}{2\dot G(p)}\dot \phi(p)\dot \phi(-p)
\]
and needs to be mapped to

\[ = \frac{1}{2 } \int_{\eps_{dS}}^{\infty} d\eta \int \Dp \left[ \left(\frac{L}{\eta}\right)^{D-1} \lbrace (\p_\eta y^{dS})^2- p^2{y^{dS}}^2-\frac{m^2 L^2}{\eta^2} {y^{dS}}^2 \rbrace \right]\]

Here $\eta = Le^{\frac{t}{L}}$. We do the field redefinition of boundary field 

\[ \phi= fy^{dS} \]

$f$ is a scale dependent quantity which is related to Green's function $G$ as $f^2=-\left(\frac{\eta}{L}\right)^{-D}\dot{G}$. Performing the same manipulations as in  \cite{Sathiapalan:2017}, one can get the constraint on f as,

\[\left(\frac{\eta}{L}\right)^{d-1} \left( \left(\frac{\eta}{L}\right)^{-d+1} \frac{d}{d\eta} \right)^2 e^{-\ln f}= \left(\frac{\eta}{L}\right)^{-d+1} \left( -p^2-\frac{m^2 L^2}{\eta^2} \right) e^{-\ln f}\]

\[ \frac{-d+1}{\eta} \frac{\p}{\p \eta} \frac{1}{f}+ \frac{\p^2}{\p \eta^2} \frac{1}{f}=\left(-p^2-\frac{m^2 L^2}{\eta^2}\right) \frac{1}{f}\]

The solutions are  $\left(\frac{\eta}{L}\right)^{d/2} H_\nu^{(1)}(p\eta)$ and $\left(\frac{\eta}{L}\right)^{d/2} H_\nu^{(2)}(p\eta)$ with $\nu^2=\frac{d^2}{4}- m^2L^2$.

\vspace{0.2 in}

The $\frac{1}{f}$ can be written in general as( note $f$ is dimensionless),

\be
\frac{1}{f(p,\eta)}= \left(\frac{\eta}{L}\right)^{d/2} \left(A H_\nu^{(1)}(p\eta)+ B H_\nu^{(2)}(p\eta)\right)
\ee

and the Green's function is \footnote{We use the term Green function by analogy with the EAdS case, where $G$ is the propagator of the boundary CFT. Also see for instance \cite{Das:2013}.} 

\[ G(p\eta)= \frac{C H_\nu^{(1)}(p\eta)+ D H_\nu^{(2)}(p\eta)}{A H_\nu^{(1)}(p\eta)+ B H_\nu^{(2)}(p\eta)}\]

Physically one can expect $G(p\eta \rightarrow \infty)=0$  which yields,

\be\label{constriant}
C H_\nu^{(1)}(p\eta)+ D H_\nu^{(2)}(p\eta)=0
\ee

The asymptotic forms of Hankel functions of both kind for large arguments are,

\[ H_\nu^{(1)}(z) \sim \sqrt{\frac{2}{\pi z}} e^{i(z-\frac{\nu\pi}{2}-\frac{\pi}{4})}~~~~ -\pi<arg~z< 2\pi  \]
\[ H_\nu^{(2)}(z) \sim \sqrt{\frac{2}{\pi z}} e^{-i(z-\frac{\nu\pi}{2}-\frac{\pi}{4})}~~~~ -2\pi<arg~z< \pi \]

The presence of the oscillatory functions will not let eq.\ref{constriant} to be satisfied. Hence we analytically continue the argument of Green's function G. The choice of direction of the analytic continuation is based on the anticipation that the bulk field will have positive frequency. Hence we take

\be\label{analytic}
\eta=-iz
\ee

which prompts us to make $C=0$. Also, from the constraint $AD-BC=1$ we get $A=\frac{1}{D}$.

Hence the Green's function now takes the form,

\[ G(pz)= \frac{D H_\nu^{(2)}(ipz)}{\frac{1}{D} H_\nu^{(1)}(ipz)+ B H_\nu^{(2)}(ipz)} \]

Next another constraint will come from the fact that boundary Green's function is $\gamma_{dS}~ p^{-2\nu}$. So in the limit of $z \rightarrow 0$ using the formulae,

\[ H_\nu^{(1)}(z)=i Y_\nu(z);~ H_\nu^{(2)}(z)=-iY_\nu(z);~ Y_\nu(z)= -\frac{\Gamma(\nu)}{\pi} \left(\frac{2}{z}\right)^\nu\]

One can get,

\[ \frac{-iD}{\frac{i}{D}-iB}= \gamma_{dS}~ p^{-2\nu} \]

On the other side, $f$ should become a p independent constant at boundary $x=0$ so that it does not modify the boundary Green's function, also $y^{dS}$ and $f$ should become same field in boundary field theory. This gives,

\[ \frac{i}{D}-iB=p^\nu \]

Finally we get,

\[ D=i\gamma_{dS}~ p^{-\nu}~;~B=i\left(1-\frac{1}{\gamma_{dS}}\right)p^\nu\] 

The bulk field $y^{dS}$ is given by, 

\[ y^{dS} = b_{dS} \frac{G}{f}= b_{dS} (i\gamma p^{-\nu})\frac{1}{L^{d/2}}x^{d/2}H_\nu^{(2)}(ipx) \]

If we analytically continue back to $\eta$ we get,

\[ y^{dS}= b_{dS} (i\gamma p^{-\nu})\frac{1}{L^{d/2}}(-i\eta)^{d/2}H_\nu^{(2)}(p\eta) \]

If the field $y^{dS}$ satisfies $y_0^{dS}$ at $\eta=\epsilon_{dS}$ then,

\begin{equation}\label{dsbulk}
y^{dS}= y_0^{dS}\frac{\eta^{d/2}}{\eps_{dS}^{d/2}} \frac{H_\nu ^{(2)}(p\eta)}{H_\nu ^{(2)}(p\eps_{dS})}
\end{equation}

$y_{dS}$ satisfies Bunch-Davies condition.

\paragraph{Relation between bulk fields in EAdS and dS}

The bulk field in EAdS space is given by,

\be
y^{EAdS}=y_0^{EAdS} \frac{z^{d/2}}{\eps^{d/2}} \frac{K_\alpha(pz)}{K_\alpha(p\eps)}
\ee
Let's apply the analytic continuation continuation $i\eta=z$ and $iL=R$. First of all, $\alpha$ becomes $\nu$. $\eps$ is replaced by $i\eps$. We get,
\be
y^{EAdS}|_{z= i\eta,~R= iL}= y_0^{EAdS}|_{z=i\eta,~R=iL} \frac{(i\eta)^{d/2}}{(i\eps)^{d/2}} \frac{K_\nu(ip\eta)}{K_\nu(ip\eps)}
\ee
As,
\be
y_0^{EAdS}= b_{EAdS}~\eps_{EAdS}^{d/2} \frac{\gamma_{EAdS}~K_\alpha(p\eps)}{p^\alpha}
\ee

Using the relation between $K_\alpha(x)$ and $H_\alpha(x)$,

\begin{equation} 
	\begin{split}
	K_{\alpha}(x) & = \frac{\pi}{2} i^{\alpha+1} H_\alpha^{(1)}(ix); ~ -\pi < arg~ x \leq \frac{\pi}{2}\\
	 & = \frac{\pi}{2}(-i)^{\alpha+1} H_\alpha^{(2)}(-ix); ~ -\frac{\pi}{2} < arg~ x \leq \ \pi  
	\end{split}
	\end{equation}
	
Here also we want to ensure the bulk field to be of positive frequency, hence choosing $H^{(2)}(x)$.

\[y_0^{EAdS}|_{z=i\eta,~R=iL}= \frac{\pi}{2} (i)^{d/2+\alpha+1} b_{EAdS} \eps^{d/2} \gamma_{EAdS} \frac{H_\alpha^{(2)}(p\eps)}{p^\alpha}\]
\[= \frac{b_{EAdS}}{b_{dS}}\frac{\gamma_{EAdS}}{\gamma_{dS}}\frac{\pi}{2} (i)^{d/2+\alpha+1} y_0^{dS}\]

Hence,

\begin{align}
\nonumber y_{EAdS}|_{z=i\eta,~R=iL} = &\frac{b_{EAdS}}{b_{dS}}\frac{\gamma_{EAdS}}{\gamma_{dS}}\frac{\pi}{2} (i)^{d/2+\alpha+1} y_0^{dS} \frac{\eta^{d/2}}{\eps^{d/2}}\frac{H_\alpha^{(2)}(p\eta)}{H_\alpha^{(2)}(p\eps)}\\
&= \frac{b_{EAdS}}{b_{dS}}\frac{\gamma_{EAdS}}{\gamma_{dS}}\frac{\pi}{2} (i)^{d/2+\alpha+1} y_{dS}
\end{align}

Upto various normalization constants we see that they agree.

\section{Summary and Conclusions}
\label{conclusion}
In \cite{Sathiapalan:2017, Sathiapalan:2019} an evolution operator for an ERG equation of a perturbed $D$-dimensional free field theory in flat space was mapped to a field theory action in $AdS_{D+1}$. Similar mappings were done subsequently for the interacting $O(N)$ model at both the free fixed point and at the Wilson-Fisher fixed point \cite{Sathiapalan:2020}. The main aim of this paper was to understand better the mapping used in these papers and to see if there are other examples. A related question was that of analytic continuation of these theories. These questions can posed, both for the ERG equation and its evolution operator. 

It was shown that a mapping of this type can map the 
ERG evolution operator of a (zero-dimensional) field theory to the action of  a Euclidean harmonic oscillator. Furthermore the analytic continuation of the ERG evolution operator action gives the path integral for a free particle with a time dependent mass. A similar mapping takes this to a harmonic oscillator. This method also gives new way of obtaining the Ermakov-Lewis invariants for the original theory. 

The analytically continued ERG equation is a Schroedinger like equation for a free field theory wave functional. This gets mapped to the Schroedinger equation for a wave functional of a free field theory in de Sitter space.  These are summarized in Figures 1,2. This is one version of the dS-CFT correspondence. From this point of view, the QM evolution of dS field theory is also an ERG evolution of a field theory, but accompanied by an analytic continuation. An example was worked out to illustrate this correspondence.

To understand these issues further it would be useful to apply these techniques to the $O(N)$ model 
ERG equation written in \cite{Sathiapalan:2020}. This ERG equation has extra terms and thus the theory naturally has interaction terms  in the EAdS bulk action.

Similarly it would be interesting to study the connection between bulk Green functions and the QM correlation functions on the space-like time slice of these theories, as considered originally in \cite{Witten:2001,Strominger:2001,Maldacena:2002}.

\vspace{0.2 in}

\paragraph{Acknowledgements}

SD would like to thank IMSc,Chennai where part of the work was done.


\end{document}